# Damping in Ru/Co-based multilayer films with large Dzyaloshinskii-Moriya interaction

T. Ščepka[a,*], P. Neilinger[b], A. S. Samardak[c,d], A. G. Kolesnikov[c], A. V. Ognev[c], A. V. Sadovnikov[e,f], V. A. Gubanov[e], S. A. Nikitov[e,f], K. Palotás[g,h,i], E. Simon[g], L. Szunyogh[g,j], J. Dérer[a], V. Cambel[a], and M. Mruczkiewicz[a]

[a]*Institute of Electrical Engineering, Slovak Academy of Sciences, Dúbravská cesta 9, 841 04 Bratislava, Slovakia*
[b]*Department of Experimental Physics, Comenius University, 842 48 Bratislava, Slovakia*
[c]*School of Natural Sciences, Far Eastern Federal University, 8 Sukhanova Str., 690950 Vladivostok, Russia*
[d]*National Research South Ural State University, 76 Lenin Prospect, 454080 Chelyabinsk, Russia*
[e]*Laboratory "Metamaterials," Saratov State University, 83 Astrakhanskaya Str., 410012 Saratov, Russia*
[f]*Kotel'nikov Institute of Radioengineering and Electronics, Russian Academy of Sciences, 11 Mokhovaya Str.,125009 Moscow, Russia*
[g]*Department of Theoretical Physics, Budapest University of Technology and Economics, H-1111 Budapest, Hungary*
[h]*Wigner Research Center for Physics, Hungarian Academy of Sciences, H-1525 Budapest, Hungary*
[i]*MTA-SZTE Reaction Kinetics and Surface Chemistry Research Group, University of Szeged, H-6720 Szeged, Hungary*
[j]*MTA-BME Condensed Matter Research Group, Budapest University of Technology and Economics, Budafoki út 8., H-1111 Budapest, Hungary*

**ABSTRACT**

Recent development of the magnetic material engineering led to achievement of the systems with a high interfacial Dzyaloshinskii-Moriya interaction (DMI). As a result, the formation of non-collinear magnetic soliton states or nonreciprocal spin wave dynamics is achievable. Typically used materials are based on bi-layers Heavy Metal/Ferromagnet, e.g., Pt/Co. These layers are characterized not only by a strong DMI, but also by the spin pumping effect and the resulting relatively large damping. Here, we show that the considerable interfacial DMI can be also present in bi-layers based on Ru/Co, characterized with low spin pumping effect and low damping. It is therefore a good candidate for the dynamical studies and implementations of chiral DMI. It is demonstrated by theoretical calculations that the value of DMI can be strongly affected and controlled by the strain of the lattice. We show a systematic experimental and theoretical comparison of magnetic material parameters between Pt/Co and Ru/Co bi-layers as a deserving candidate for spintronic and spin-orbitronic applications.

[*]tomas.scepka@savba.sk

## I. INTRODUCTION

Demand for further improvements in the speed and storage capacity of magnetic storage devices and decrease of their power consumption leads to continued interest in spintronic phenomena such as the spin transfer torque [1], [2], spin-orbit torques [3], and interfacial Dzyaloshinskii–Moriya interaction (IDMI) [4]. The spin-orbit interaction plays an important role in these effects, as it couples the spin and charge degrees of freedom of an electron, which enables the conversion of an electric current into a transverse spin current and vice versa.

The IDMI can be induced by the large spin-orbit coupling of a heavy metal (HM) in contact with ultrathin ferromagnetic material (FM) where the inversion symmetry is broken at the surface. If the IDMI is strong enough, it can change the static as well as the dynamic properties of the system. Indeed, it converts the magnetostatically favorable Bloch wall into a chiral Néel wall and induces chiral canting of spins which leads to stabilization of noncolinear magnetic states, e.g., magnetic skyrmions [5], radial vortices, stripe- or labyrinth-like domains. The IDMI and chiral canting leads also to nonreciprocal spin wave dynamics [6], [7], [8]. Taking advantage of the fast motion of chiral textures and nonreciprocal dynamics can satisfy the demands for high-density data storage, low power consumption and high processing speed.

Recently, the promising solutions for a formation of non-collinear spin configurations are trilayer structures $HM_1/FM/HM_2$, where $HM_1$ and $HM_2$ are bottom and top layers of heavy metals, respectively, with the enhanced spin-orbit effects, including the IDMI. The IDMI is usually characterized by its effective ($D_{eff}$) or surface ($D_S$) constants. The precise determination of the sign and value of the IDMI constant is interesting for both fundamental research and applications. It is possible to tune the total IDMI constant by choosing suitable $HM_1/FM$ and $FM/HM_2$ interfaces. Since IDMI arising on the top and bottom interface can have the same or opposite sign, either enhancement or cancelation of the total IDMI constant can appear. It is also known that quasi-symmetrical trilayers, like Pt/Co/Pt, possess a non-zero IDMI due to the different quality of the top and bottom interface [9]. Typical material systems used for high IDMI and perpendicular magnetic anisotropy (PMA) are Co-based multilayers, specifically composed of a Pt/Co bilayer, such as: Pt/Co/AlOx [10], [11], Pt/Co/MgO [12], Pt/Co/Ta [13], Pt/Co/(W, Ta, Pd) [14], Pt/Co-Ni/Ta [15], Ir/Fe/Co/Pt [16], Pt/Co/Cu/AlOx [17], Pt/Co/Os/Pt [18], Pt/Cu/Co/Pt [19]. Even though Pt/Co-based structures reveal relatively high IDMI constants, at the same time they suffer from a large magnetic damping [20]. The damping constant is also an important magnetic parameter as it determines the magnetization dynamics, such as the switching rate of magnetization reversal, domain wall velocity and spin wave propagation. The extreme magnetic damping present due to spin pumping mechanism in Pt/Co($t_{Co}$ ~ 1 nm) systems makes dynamical studies practically impossible.

In our approach, we studied Ru/Co/Cu/Ru structures, where we focused on spin-orbit coupling (SOC)-induced interfacial effects arising mainly from the bottom $HM_1/FM$ interface, while a Co/Cu interface is considered as a negligible contribution in both the magnetic damping and the IDMI evaluation [21], [22]. Moreover, we initially chose the Cu layer because of its immiscibility with Co, which was previously published to prevent Co/Pt intermixing at the upper interface or in a multilayer stack [23]. A few works showed that Ru/Co-based structures can have a significant IDMI constant [24], [25], [26]. Taking into consideration an interplay between Heisenberg exchange, dipole, and anisotropy interaction, it can be above the critical value of IDMI to stabilize the non-collinear states [27].

The magnetic damping in these structures has not been well studied yet. In Ref. [28] a damping comparison between Pt/[Co/Ni] and Ru/[Co/Ni] multilayers revealed very similar values of the Gilbert damping constant. The damping constant of a Ru/[Co/Ni] structure was also kept after post-annealing up to 400°C. In Ref. [22] it was shown experimentally that Pt/Co/Ir and Pt/Co/Ru multilayer films can have similar IDMI values. It is also known that if a Ru spacer layer is sandwiched between two Co layers, it will produce an interlayer exchange coupling (IEC). The large IEC may provide an additional magnetic energy to tune non-collinear states within the superlattice structures. We propose Ru/Co-based structure as a candidate for material systems that can host a high enough IDMI constant and lower magnetic damping compared to Pt/Co, Pt/Py, or Pt/CoFeB-based ones. Such structures could support long propagation of spin waves, dynamical modes of skyrmions, etc.

The structural and interfacial properties were studied by X-ray reflectivity (XRR) measurement technique at CuKα radiation with a wavelength of 1.54 Å. Magnetic properties of the films were investigated with vibrating sample magnetometer (VSM) at room temperature. The ferromagnetic resonance (FMR) measurements were carried out at room temperature using a coplanar waveguide (CPW) with a broad frequency range up to 25 GHz. Brillouin light scattering (BLS) spectroscopy measurements were performed in the Damon-Eshbach (DE) geometry in order to define the sign and value of IDMI. First-principles (*ab initio*) calculations and atomic spin-model simulations were performed to explore competing exchange interactions in ultrathin film systems.

## II.     EXPERIMENTAL AND COMPUTATIONAL DETAILS

All polycrystalline films presented here were deposited on a silicon wafer by both direct current and rf magnetron sputtering in a UHV system at room temperature. Argon gas (~$2 \times 10^{-3}$ mbar) was used during the sputtering process with a background pressure between $1 \times 10^{-7}$ and $10^{-8}$ mbar. The full film stack was as following, Si/SiO$_2$ substrate/Ta(4)/Pt(3) or Ru(3)/Co($t_{Co}$)/Cu(4)/Pt or Ru, where the numbers in parentheses represent the layer thickness in nm. The nominal Co thickness $t_{Co}$ was standardly set to 1, 2, or 3 nm. The lowermost Ta layer is a seed layer used to enhance the crystallinity of the films, and the uppermost Pt or Ru layer is a protective layer used to prevent oxidation.

The in-plane (IP) and out-of-plane (OOP) magnetic hysteresis loops were measured by VSM and all samples studied revealed an IP magnetic anisotropy.

A broadband FMR spectrometer based on the vector-network analyzer (VNA-FMR) technique was used to obtain the FMR absorption profiles of multilayers in the IP geometry with an external magnetic field applied in the same plane. An IP microwave field with a frequency up to 25 GHz was applied to the sample using a grounded CPW with a 240 µm wide central stripe. The samples were placed face down on the CPW so that the substrate was always the furthest layer in the investigated multilayer structures. The drive-field frequency is varied at a fixed static field and the FMR absorption profile is extracted from standard $S_{21}$ parameter measurements of the transmission line. The FMR spectra are measured at different magnetic fields to determine the magnetic damping constant $\alpha$.

The BLS, which detects light inelastically scattered by excitations, is currently a suitable technique for evaluating the DMI parameter. The influence of IDMI on the spin wave (SW) spectrum is well-known, both theoretically [7] and experimentally [10], [29], [11]. The IDMI induces a characteristic non-reciprocity of the oppositely propagating SWs, resulting to different frequency values for the same wave-number. The Damon-Eshbach (DE) configuration is used, since the frequency shift of the two counter-propagating SWs is the largest. The BLS experiments were performed in the back-scattering geometry and the spectra were obtained by accumulating photons typically for 10-30 hours. This configuration allows to determine the frequency position of the BLS peaks with a precision around 0.1 GHz. The Stokes ($f_S$) and anti-Stokes ($f_{AS}$) frequencies were determined from Lorentzian fits to the spectral peaks.

Using the relativistic screened Korringa-Kohn-Rostoker (SKKR) method [30], [31], [32] the magnetic properties of Co layers of various thicknesses sandwiched between Ru, Cu and Pt, Cu metallic layers were investigated. The local spin-density approximation of density functional theory (DFT) as parameterized by Vosko et al. [33] and the atomic sphere approximation with an angular momentum cut-off of $l_{max} = 2$ were used. The magnetic interfaces, consisting of 6 atomic layers Ru(Pt), n = 1…6 Co layers, 6 layers Cu, and 6 layers Ru(Pt) between semi-infinite Ru(Pt) substrates, were treated self-consistently. For modeling the geometries of the interfaces the experimental IP lattice constant ($a_{2D}$) of the Ru(0001) and Pt(111) surface was used, respectively, and hcp layer growth was assumed for Ru and Co layers, while for Pt and Cu layers fcc growth was considered. The distances between the atomic layers were optimized in terms of VASP calculations [34], [35]. Inward layer relaxations between 5% and 15% for the Co atomic layers were found depending of their thickness. The interlayer distances at the interfaces (Ru-Co, Co-Cu; Pt-Co, Co-Cu) were also optimized, and they were taken into account in the self-consistent SKKR calculations.

In order to investigate the magnetic structure of the Co layers in the above-described interfaces a generalized classical Heisenberg model was used following the sign conventions of Ref. [36]. The magnetic exchange coupling tensor was determined in terms of the relativistic torque method, based on calculating the energy costs due to infinitesimal rotations of the spins at selected sites with respect to the ferromagnetic state oriented along different crystallographic directions [37], [38]. The magnetic interaction tensors were determined for all pairs of atomic spins up to a maximal distance of $5a_{2D}$, for a total of 90 neighbors including symmetrically equivalent ones [36], [39]. The calculated atomic magnetic interaction parameters for many different neighbors were transformed to effective nearest neighbor interactions following the method described in Ref. [39]. This enables a direct comparison with experimentally determined magnetic parameters for the magnetocrystalline anisotropy ($K_{eff}$) and the scalar DMI ($D_{eff}$). The ground states of the corresponding systems were also determined by using atomistic spin-dynamics simulations.

### III. RESULTS AND DISCUSSION

After the sputtering process the cross-sectioned sample was firstly imaged in the high-angle annular dark-field (HAADF) mode of a high resolution transmission electron microscope (HR-TEM). This is shown in Figure 1 for a 10 times repeated Pt(3)/Co(1)/Au(3) multilayer sample. HAADF imaging is sensitive to atomic number ($Z$) of the material so images with low $Z$ appear darker than those with high $Z$ [40]. There is no significant contrast between Pt and Au layers since they have very similar $Z$. The continuous Co layers with the thickness of 1 nm are clearly visible in the image. We assumed that similar structure is for all samples presented here, including Ru/Co-based samples.

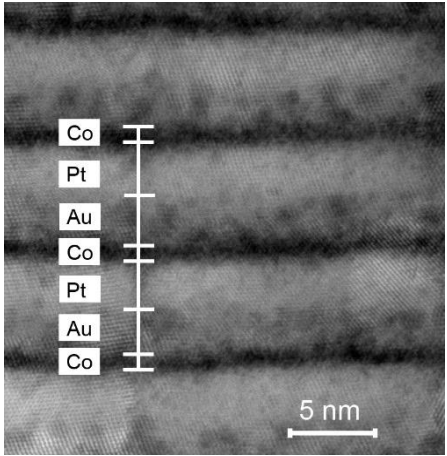

**Figure 1** The HR-TEM cross-section image. Multilayer stack of [Pt(3)/Co(1)/Au(3)]$_{x10}$ showing the continuous Co layers with the thickness of 1 nm. Marks indicating the different layers are superimposed as a guide to the eye.

*PPMS magnetometry and Effective magnetic anisotropy energy*

To extract magnetic parameters of the thin films, we carried out magnetic characterization of Ru/Co($t_{Co}$)/Cu/Ru samples by using a VSM PPMS magnetometer. The normalized magnetic hysteresis loops of the two multilayers ($t_{Co}$ = 1 and 3 nm) at room temperature are shown in Figure 2. An IP magnetic anisotropy is dominant for both samples. The saturation magnetization $M_S$ approximately corresponding to 950 and 930 kA/m, the coercivity field of 1.2 and 1.6 mT, and anisotropy field of 0.65 and 1.3 T for Ru/Co(1)/Cu/Ru and Ru/Co(3)/Cu/Ru were measured, respectively. Those values of $M_S$ are smaller than the bulk magnetization density of fcc Cobalt 1400 kA/m, probably due to dead layers arising from intermixing between a heavy metal and Co. Dead layer thickness in our Ru/Co films is estimated to be around 0.2 nm, similarly as in Ref. [41].

Using XRR spectra, we deduced the thickness and density of each layer as well as the roughness of interfaces. In the case when an extracted value of thickness was close to the nominal value, we used the nominal value within the paper. From the fitting of the XRR data we obtained the root mean square roughness of the most interfaces between 0.25 and 0.30 nm.

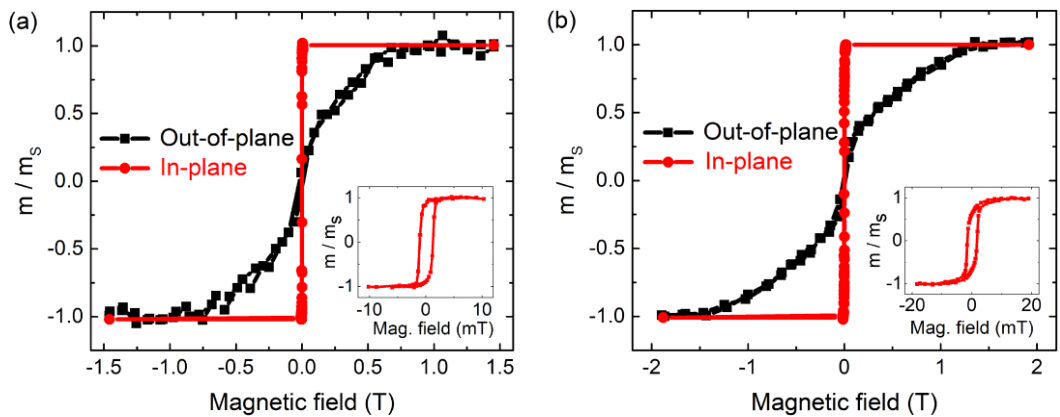

**Figure 2** Room-temperature magnetic hysteresis loops measured for (a) Ru/Co(1)/Cu/Ru and (b) Ru/Co(3)/Cu/Ru samples along the IP and OOP directions, respectively. Each loop is normalized to the saturated magnetic moment. The insets show a zoom of the IP measurement.

The critical Co layer thickness was determined in Ru/Co($t_{Co}$)/Cu/Ru structures from the variation of the anisotropy field as a function of thickness. Figure 3 shows the variation of $K_{eff} \times t_{Co}$ as a function of Co thickness. The vertical axis intercept equals twice the surface anisotropy ($K_s$), whereas the slope gives the volume contribution. Based on our measurements, the critical Co thickness is only slightly depending on the Ru layer thickness lying underneath. For $t_{Co} > 0.7$ nm, the effective anisotropy is negative such that the easy axis of the samples is in-plane.

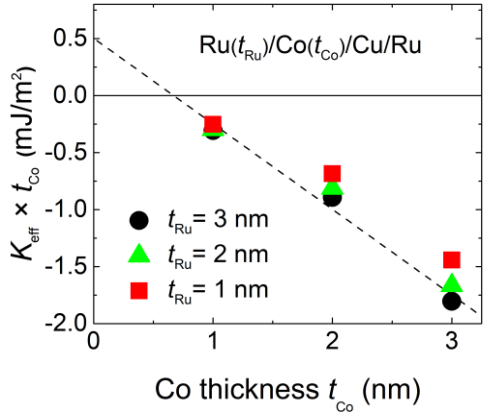

**Figure 3** Effective magnetic anisotropy. $K_{eff} \times t_{Co}$ versus $t_{Co}$ plot for Ru($t_{Ru}$)/Co($t_{Co}$)/Cu/Ru samples. For $t_{Co} \sim 0.7$ nm, the effective uniaxial anisotropy changes from positive to negative values, which means the direction of the easy axis changes from perpendicular to the in-plane. The dashed line represents a linear fit of the multilayer with $t_{Ru} = 3$ nm.

Here, it should be mentioned that all Ru/Co-based samples in our study have IP easy axis, because we use rather thick Co (nominally $t_{Co} \geq 1$ nm) layers. When the ferromagnetic layer is too thin ($t_{Co} < 0.6$ nm), both the magnetic damping and spin-wave non-reciprocity evaluation is not accessible by our method. This is the reason why we limited our measurements only on the multilayer films with $t_{Co} \geq 1$ nm.

*Ferromagnetic resonance spectroscopy*

The FMR frequency $f$ and linewidth $\Delta H$ were extracted from the measured transmission spectra as described in Ref. [42]. The FMR resonance condition for a FM thin film in IP geometry is given by the Kittel formula [43],

$$f = \frac{\gamma \mu_0}{2\pi} \sqrt{H_{res}(H_{res}+M_{eff})}, \tag{1}$$

where $H_{res}$ is the applied magnetic field at resonance, $M_{eff}$ is the effective magnetization, and $\gamma = g\mu_B/\hbar$ is the gyromagnetic ratio which is proportional to the Landé factor $g$.

The FMR frequency is varied at a fixed static field and the absorption profile is extracted from the standard $S_{21}$ parameter measurements of the transmission line. The dependence of the FMR frequency on the applied magnetic field for the Ru/Co(3)/Cu/Ru film is shown in Figure 4. The absorption amplitude, which is a measure of the FMR signal strength, decreases with the Co layer thickness and, hand in hand with an increase of the damping, makes the measurement of thin layers challenging. The spectra were measured multiple times in order to increase the signal-to-noise ratio and a non-resonant background is subtracted by a polynomial fit. Even though it is not discussed here, a possible solution is to utilize high stacking numbers, where a strong enhancement of the FMR signal can be obtained without a significant change in the magnetic properties of the layers [42].

The resonant field and the linewidth were determined from Lorentzian fits of the transmission spectra. The Landé *g*-factor and effective magnetization were obtained from the fit of the resonant frequency dependence on the resonant field.

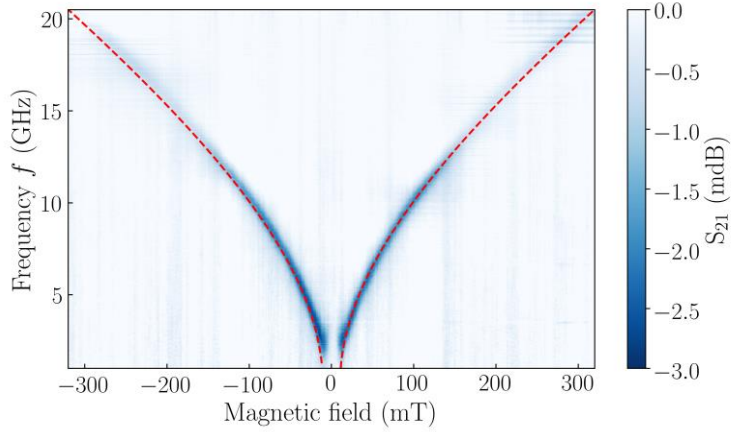

**Figure 4** Color-coded absorption spectra of the Ru/Co(3)/Cu/Ru sample and the fit (red dashed lines) of the resonant field $H_{res}$ to the Kittel formula.

The intrinsic magnetic damping *α* in a dimensionless form can be estimated from the frequency dependence of the half-width at half-maximum $\Delta H$ of the resonant absorption as [44], [45]:

$$\mu_0 \Delta H = \mu_0 \Delta H_0 + \frac{4\pi\alpha}{\gamma} f, \tag{2}$$

where the $\Delta H_0$ is the inhomogeneous zero-frequency broadening that contributes to the FMR linewidth. It is expected that inhomogeneity in our samples may come from a wide variety of origins including lattice defects and magnetic impurities, nonuniform stresses, or from surface anisotropy with film thickness random variations [46].

In Figure 5, total magnetic damping is depicted for different multilayer structures. The enhancement of the damping for the ultrathin ferromagnetic Pt/Co-based films can be attributed to the spin pumping mechanism. This effect is significantly suppressed in the case of Ru/Co-based structures. By introducing a tantalum buffer layer, that decreases the interfacial roughness, even slightly lower damping can be achieved for the thinnest Co multilayer film [47].

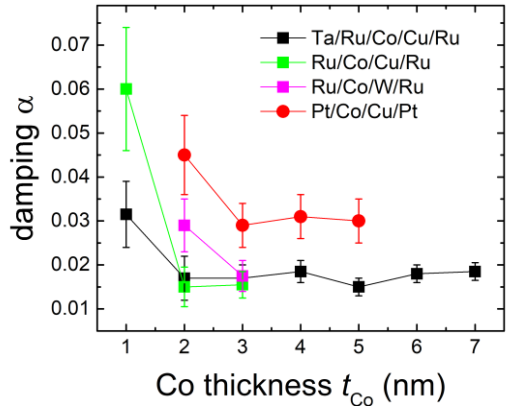

**Figure 5** The magnetic damping versus the Co layer thickness comparison in Pt/Co and Ru/Co-based samples.

For Co thicknesses ≥ 3 nm, the value of magnetic damping in Pt/Co-based and Ru/Co-based samples reaches the constant of 0.03 and 0.017, respectively. For thinner Co thicknesses, the magnetic damping in Pt/Co-based samples is rapidly increasing up to a value where it is not possible to evaluate by our technique. Instead, the magnetic damping in Ru/Co-based sample with the thinnest Co layer is around 0.03, what is comparable to the value of magnetic damping for large Co thicknesses in Pt/Co-based samples.

*Brillouin light scattering*

In this study, we investigated propagating spin wave dispersion by performing a BLS spectroscopy. As a result, IDMI energy density can be extracted from the two different SW frequencies (Stokes and anti-Stokes). In the systems with no IDMI, the SW frequencies of Stokes and anti-Stokes peaks are at the same position or slightly different due to the PMA energy difference between top and bottom interfaces of the ferromagnet. However, the propagating SWs with opposite wavevectors ($\pm k$) can be significantly different due to IDMI interaction and the resulting frequency difference ($\Delta f$) is measurable by the BLS technique. Here, the DC external magnetic field was applied parallel to the film surface and perpendicular to the scattering plane. We used an applied magnetic field ($H = 0.4 - 1$ T) with the fixed $k_x = 0.011$ nm$^{-1}$ for magnetic field dependence and an incident angle of light $\theta$ (15° - 80°) corresponding to $k_x = 0.006 - 0.023$ nm$^{-1}$ with the fixed magnetic field of 0.2 T for dispersion relation measurements.

The typical BLS spectra of a Ru/Co(3)/Cu/Ru sample is shown in Figure 6. The red solid line represents experimentally measured data and the black dashed line is a mirrored curve for a direct comparison. The green dashed vertical lines illustrate the SW frequency obtained from Lorentzian fit and green arrows indicate the $\Delta f$, in this case it is approximately 0.5 GHz. The GHz range of the $\Delta f$ is a first signature of the IDMI interaction. The data accumulation time for each spectrum was about 600 min.

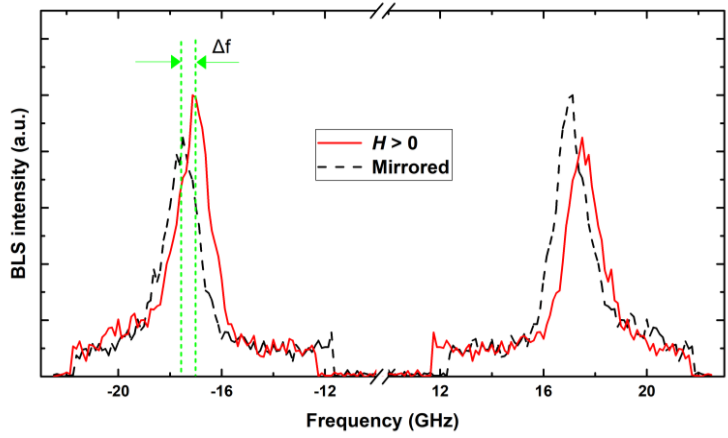

**Figure 6** The BLS spectrum recorded for Ru/Co(3)/Cu/Ru sample with an applied magnetic field $H_{ext}$ = 0.22 T (red curves) and their mirrored curves drawn as a black dashed line. The incident angle of light was fixed at $\theta = 15°$ ($k = 0.006$ nm$^{-1}$).

The shift between the Stokes and anti-Stokes frequencies of spin waves of a given wavelength is directly proportional to the DMI value as [10], [29]:

$$\Delta f = f_S - f_{AS} = \frac{2\gamma}{\pi M_S} D_{eff}\, k, \qquad (3)$$

where, $f_S$, $f_{AS}$, $\gamma$, $M_s$, $k = 4\pi\sin\theta/\lambda$, and $D_{eff}$ are Stokes, anti-Stokes SW frequencies, the gyromagnetic ratio, the saturation magnetization, the SW in-plane wave vector, and the effective IDMI constant, respectively. $D_{eff} = D_S/t_{FM}$, where $D_S$ is the surface IDMI constant and $t_{FM}$ is the ferromagnetic layer thickness. From Eq. 2 one can see that the $\Delta f$ is invariant on an applied magnetic field, thus the $\Delta f$ uncertainties can be minimized by the field-dependent measurements. The measured SW frequencies as a function of the magnetic field for a Ru/Co(3)/Cu/Ru sample are depicted in Fig. 7, where one can see that the $\Delta f$ is nearly a constant with the average value of 0.85 GHz. The inset shows the $\Delta f$ fluctuations as a function of an applied magnetic field, where nearly all experimental points are within the BLS detection limit (~ 125 MHz). From the previous magnetic measurements it is obvious that fields ≥ 0.4 T are high enough to saturate IP magnetization. The $\Delta f$ value for specific samples was measured as a function of an IP applied magnetic field and varying the incident angle of the probing light.

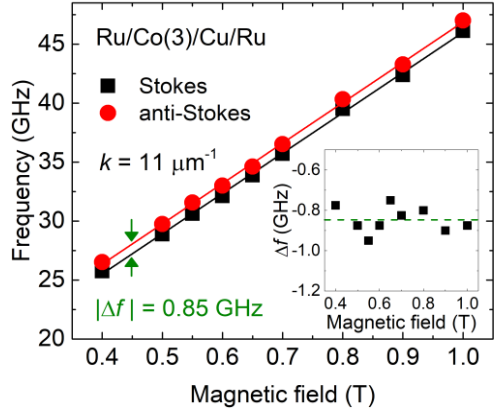

**Figure 7** The BLS measurements of Ru/Co(3)/Cu/Ru sample. The SW frequencies as a function of applied magnetic field for $k = 0.011$ nm$^{-1}$. The frequency difference between Stokes and anti-Stokes peaks is constant in the range of 0.4 – 1 T. Symbols represent the experimental data points and the lines are their linear fits. The inset shows the $\Delta f$ as a function of an applied magnetic field. The dashed green line represents the averaged value of 0.85 GHz.

In the next step, the dispersion relation of SWs was measured for the samples with the Co thickness of 1 and 3 nm. The dependences of the frequency on the wave vector for two various thicknesses are plotted in Fig. 8a. It is evident that at finite $k$, the spin-wave frequencies propagating along two opposite directions differ significantly and this asymmetry increases with increase in the value of $k$. Even though our BLS setup is capable to measure in a limited region from 0.006 to 0.023 nm$^{-1}$, one can see the parabolic SW dispersion relations with minima shifted from $k = 0$. This asymmetry in dispersion relations is a clear evidence of the IDMI interaction in our samples. Fig. 8b shows the $\Delta f$ versus $k$ for two opposite directions of an applied magnetic field for $t_{Co} = 3$nm. The linear fits of the experimental data reveal nearly the same tendency. All deviations of the experimental points are mainly caused by the BLS detection limit and determination of the SW peak positions. Importantly, if $f_S$ is lower than $f_{AS}$, than the resulting IDMI constant is negative for positive applied magnetic field and vice versa. The IDMI sign of Ru/Co samples was found to be the same (negative) as that of Pt/Co. Thus, using the ternary superlattices based on Pt/Co/Ru is definitely not suitable in terms of the IDMI enhancement.

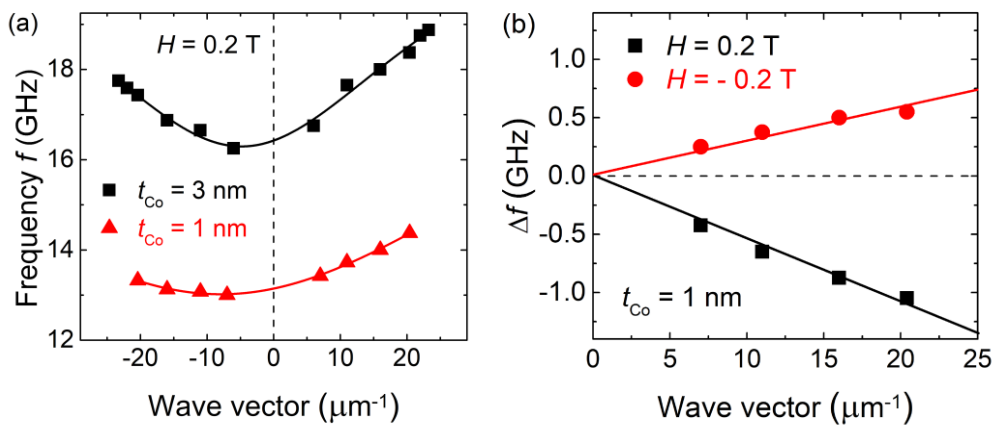

**Figure 8** (a) The asymmetric dispersion relation measured by the BLS for two Ru/Co(1 or 3)/Cu/Ru samples. The curves are the Lorentzian fits of the experimental points and the dashed line represents $k = 0$. (b) Frequency shift $\Delta f$ depending on the wave vector $k$ at the positive ($H = 0.2$ T) and negative ($H = -0.2$ T) applied field. The lines are linear approximation of the experimental data with the origin intersection (0, 0) and the dashed line separates the positive and negative values of the frequency shift.

In multilayer systems like those discussed here, the IDMI is expected to be an interface-like effect, where its effective strength scales with the inverse of the ferromagnetic layer thickness. In the case of the Pt/Co-based structures, the tendency and the IDMI values are similar to previously reported studies [11]. However, for Ru/Co-based multilayers, the effective IDMI coefficient is not decreasing with the inversed Co thickness. The reduction of the IDMI for $t_{Co} = 1$ nm in Ru/Co-based system may be attributed to the different strain conditions at the interface between Ru and Co. Using Eq. 2 we can declare that values of the frequency shift in GHz units correspond to values of the effective IDMI parameter in mJ/m$^2$. It means that the $D_{eff}$(Ru/Co) is halved in the case of $t_{Co} = 1$ nm, while $D_{eff}$(Ru/Co) $\simeq D_{eff}$(Pt/Co) for 3 nm-thick Co layer (Tab. 1). The value of $D_{eff}$ for Ru/Co($t_{Co} = 3$ nm) is surprisingly high and it does not match the theoretical predictions. Such a strong non-reciprocity in our thicker sample is still not well explained by our experiments.

**Table 1** The frequency shift and the effective IDMI constants extracted for Pt/Co and Ru/Co-based samples with two thicknesses of the ferromagnetic layer.

|  | $\Delta f$ (GHz) ($t_{Co} = 1$ nm) | $D_{eff}$ (mJ.m$^{-2}$) ($t_{Co} = 1$ nm) | $\Delta f$ (GHz) ($t_{Co} = 3$ nm) | $D_{eff}$ (mJ.m$^{-2}$) ($t_{Co} = 3$ nm) |
|---|---|---|---|---|
| Pt/Co/Cu/Pt | -1.23 | ~ -1.3 | -0.89 | ~ -0.87 |
| Ru/Co/Cu/Ru | -0.65 | ~ -0.63 | -0.85 | ~ -0.83 |

*Theoretical results*

Fig. 9 shows the calculated effective nearest neighbor magnetic interactions of Co layers for all of the considered Ru/Co$_n$/Cu$_6$/Ru and Pt/Co$_n$/Cu$_6$/Pt systems (subscript here and in this section means the number of atomic layers; n = 1...6, i.e. $t_{Co} \leq 1$ nm). Note that in this section the quantities with "eff" index correspond to calculated results, and $D_{eff}$ and $K_{eff}$ are not equal to identically denoted quantities in the experimental sections above. As can be seen in Fig. 9, the nearest neighbor FM isotropic couplings ($J_{eff}$) are increasing in absolute value with the increased number of Co layers for Ru and Pt interface, while the effective DM couplings ($D_{eff}$) are decreasing in absolute value, and the sign of $D_{eff}$ is the same (negative) for all cases, in agreement with the above experimental findings. According to our definition [36] the negative sign of $D_{eff}$ indicates that the counterclockwise rotation of the spins is preferred. For the Ru/Co/Cu interface $D_{eff}$ is decreased in absolute value compared to Pt/Co/Cu, also in agreement with experiments.

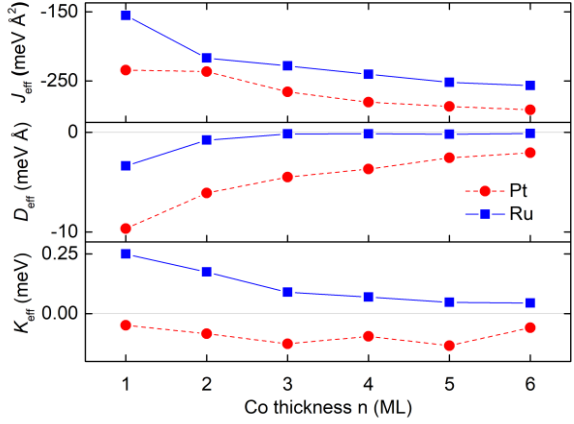

**Figure 9** Calculated effective isotropic interaction $J_{eff}$, effective DM coupling $D_{eff}$, and magnetic anisotropy $K_{eff}$, as a function of the thickness of Co layer (n) in ML unit in the Ru/Co$_n$/Cu$_6$/Ru and Pt/Co$_n$/Cu$_6$/Pt systems.

The magnetic ground state of Co was found to be ferromagnetic for all considered systems. The Co spin moments are between 1.7-1.9 (1.5-1.7) $\mu_B$ in Pt(Ru)/Co$_n$/Cu with induced moments of Pt (Ru) of about 0.3 (-0.05) $\mu_B$. The induced spin moments of Cu are even smaller (<0.02 $\mu_B$).

The magnetocrystalline anisotropy $K_{eff}$ is also shown in Fig. 9. In our definition [36] the positive and negative sign of $K_{eff}$ corresponds to the easy axis and easy-plane type of magnetic anisotropy, respectively, in line with the experimental definition. Note that the studied region for $K_{eff}$ in Fig. 9 corresponds to $t_{Co} \leq 1$ nm in Fig. 3. As Fig. 9 shows, in case of Ru/Co/Cu interfaces, out-of-plane easy axis is the preferred orientation of Co spins, while for Pt/Co/Cu interfaces easy-plane type magnetic anisotropy was observed that corresponds to in-plane spin moments of Co. This latter finding is clearly due to the Co/Cu interface since for Pt/Co/Pt out-of-plane anisotropy was observed [9], [39].

The dipole-dipole magnetic interaction on the magnetic anisotropy prefers IP orientation of the spin moments. We found that the dipole-dipole interaction is small for very thin Co layer, but it increases by increasing the number of magnetic layers [31], and this can lead to a sign change of the magnetic anisotropy in the case of Ru/Co$_6$/Cu, where IP anisotropy was experimentally observed in Fig 3 ($t_{Co} = 1$). Thus, the dipole-dipole magnetic interaction can explain the experimentally observed IP magnetic anisotropy for the thick layers $t_{Co} \geq 1$ nm.

Beyond the effective nearest neighbor magnetic interactions, the intra- and interlayer effective isotropic and DM couplings were also determined, and the calculated values are summarized in Table 2 for Ru/Co$_2$/Cu and Pt/Co$_2$/Cu interfaces. Here, the intra-layer effective isotropic couplings are much stronger ferromagnetic (negative) than the interlayer isotropic exchange; and in case of Ru/Co$_2$/Cu the interlayer coupling is slightly antiferromagnetic (positive). For both interfaces the sign of the intra-layer DM coupling in the second Co layer neighboring the Cu is reversed compared to the first Co layer neighboring the Ru or Pt. In contrast to the effective inter-layer isotropic coupling, the $D^{inter}$ is larger in absolute value than the intra-layer contributions, and the sign of $D^{inter}$ is the same as $D_1^{intra}$.

**Table 2** Calculated intra- and interlayer effective isotropic ($J$ in meVÅ$^2$ units) and DM ($D$ in meVÅ units) couplings in case of Ru/Co$_2$/Cu and Pt/Co$_2$/Cu interfaces. The layer numbering starts from the Ru(Pt) interface.

|  | $J_1^{intra}$ | $J_2^{intra}$ | $J^{inter}$ | $D_1^{intra}$ | $D_2^{intra}$ | $D^{inter}$ |
|---|---|---|---|---|---|---|
| Ru/Co$_2$/Cu | -99.72 | -125.26 | 9.09 | -0.33 | 0.87 | -1.31 |
| Pt/Co$_2$/Cu | -116.37 | -95.97 | -24.03 | -1.53 | 1.72 | -6.28 |

Further we investigate the effect of strain on the magnetic properties of Co. The above-studied Ru/Co/Cu systems were laterally confined to the IP lattice constant of the Ru substrate, 2.7059 Å, resulting in a strained Co with an enhanced IP lattice constant by 8% relative to the bulk Co. Investigating the effect of a strain-free growth of Co on its magnetic properties, the IP lattice constant of bulk Co (2.5071 Å) was also considered as a lateral confinement taking the Ru/Co$_3$/Cu system. In this model calculation the Ru was strained compressively, and the interlayer distances between atomic planes were reoptimized. The effect of the lateral confinement change 2.7059 Å (Ru) → 2.5071 Å (Co) on the magnetic properties of Co can be summarized as follows: While $J_{eff}$ (-228.1 → -188.4 meVÅ$^2$) and $K_{eff}$ (0.09 → 0.08 meV) are decreased by less than 20% in magnitude keeping their sign, $D_{eff}$ changes both sign and magnitude (-0.16 → 0.31 meVÅ), in this particular case $D_{eff}$ doubles its magnitude. This result means that the effective DM interactions in the magnetic layers are very sensitive to the atomic structural details and growth conditions of the interface, and even a chirality change of the spin rotation can be induced by strain that is indicated by the sign change of $D_{eff}$. Relevant for realistic samples, intermixing and disordered interface effects on the IDMI were also reported [9], [48].

## IV. CONCLUSION

In summary, two important parameters such as the magnetic damping and IDMI constant have been studied in $HM_1/FM/HM_2$ systems by means of the FMR and BLS technique, respectively. Here $HM_1$ is Ru or Pt, $HM_2$ is Cu, and FM is a polycrystalline Co layer of various thicknesses. Our experiments show that the magnetic damping in Pt/Co-based structures increases inversely with the Co layer thickness, and for $t_{Co} \leq 1$ nm it is not measurable by our FMR system. On the other hand, the value of the magnetic damping in Ru/Co($t_{Co}$ = 1 nm)-based multilayers is measurable. In Ru/Co($t_{Co}$ = 2 nm)-based sample, it is approximately three times lower than in the case of Pt/Co($t_{Co}$ = 2 nm). The spectroscopic measurements have showed that the IDMI value in Ru/Co($t_{Co}$ = 1 nm)-based structure is reduced by a half compared to the Pt/Co one. We can conclude that the IDMI values around 0.65 mJm$^{-2}$ are still reasonable for specific dynamical spintronic applications. Moreover, our study revealed that the sign of the IDMI for Ru/Co is matching the negative sign for Pt/Co-based samples. Theoretical calculations based on DFT explained the experimentally observed magnetic anisotropy trend and the sign of the IDMI, and provided insights into the composition of intralayer and interlayer magnetic interaction contributions. The effect of the dipole-dipole magnetic interaction on the magnetic anisotropy for thick Co layers was highlighted. It was also demonstrated that applying strain on the magnetic layers can reverse the sign of the IDMI, thus the chirality of the preferred spin rotations. Since choosing a thin film system consisting of a FM and HM layers with a low magnetic damping constant is a promising direction for efficient skyrmionic operations and nonreciprocal SW dynamics, we believe that Ru/Co-based structures are suitable candidates.

**Acknowledgments**

We gratefully acknowledge financial support from the Slovak Grant Agency APVV, grant number APVV-16-0068 (NanoSky), by the VEGA 2/0160/18, by the Era.Net RUS Plus (TSMFA), by the RFBR (#18-57-76001, #18-29-27026), by the National Research Development and Innovation Office of Hungary under projects No. K115575, No. PD120917, and No. FK124100, by the BME-Nanotechnology FIKP grant of EMMI (BME FIKP-NAT), and by the EU COST network CA17123 MAGNETOFON. A.S.S., A.G.K. and A.V.O. acknowledge the support of the Russian Foundation for Basic Research (grant 19-02-00530), the Russian Ministry of Education and Science under the state tasks (3.5178.2017/8.9 and 3.4956.2017), and the Act 211 of the Government of the Russian Federation (No. 02.A03.21.0011).